# Information and Communication Technology Reputation for XU030 Quote Companies

Sadi Evren Seker, Bilal Cankir, and Mehmet Lutfi Arslan

*Abstract*—By the increasing spread of information technology and Internet improvements, most of the large-scale companies are paying special attention to their reputation on many types of the information and communication technology. The increasing developments and penetration of new technologies into daily life, brings out paradigm shift on the perception of reputation and creates new concepts like e-societies, techno-culture and new media. Contemporary companies are trying to control their reputation over the new communities who are mostly interacting with social networks, web pages and electronic communication technologies. In this study, the reputation of top 30 Turkish companies, quoted to the Istanbul Stock Market, is studied, based on the information technology interfaces between company and society, such as social networks, blogs, wikis and web pages. The web reputation is gathered through 17 different parameters, collected from Google, Facebook, Twitter, Bing, Alexa, etc. The reputation index is calculated by z-index and f-scoring formulations after the min-max normalization of each web reputation parameter.

*Index Terms*—Business intelligence, ICT, data mining, reputation management, web-o-metric.

## I. Introduction

Based on Eurostat 2013 community survey on information and communication technology (ICT) usage and e-commerce in enterprises, the 37% of enterprises made purchase electronically (e-purchase) and 17% of enterprises made electronic sales (e-sales). From 2008 to 2012, both the e-purchase and e-sales has increased 4% and the increasing trend shows the major impact of electronic trades on the European Union (EU) [1].

Turkey is a candidate country to EU and EU is Turkey's number one import and export partner while Turkey ranks 7th in the EU's top import and 5th in export market [2].

The increasing trend on e-trade brings the concept of e-reputation, which is the reputation of company over the electronic societies through the information and communication technologies, such as social networks, wikis, blogs or web pages.

This study, should be differentiated from the web-o-metrics or cyber-metrics studies, since those studies mostly focus on the World Wide Web from the perspective of impact factors, page ranks, alt metrics or network mapping, which all of them are based on the statistics of hyperlinks [3].

In this study, the reputation of most prestigious companies in Istanbul stock market, are researched from the aspect of information and communication technology which includes the web-o-metrics and furthermore the statistics related to wikis, blogs and social media are also taken into account.

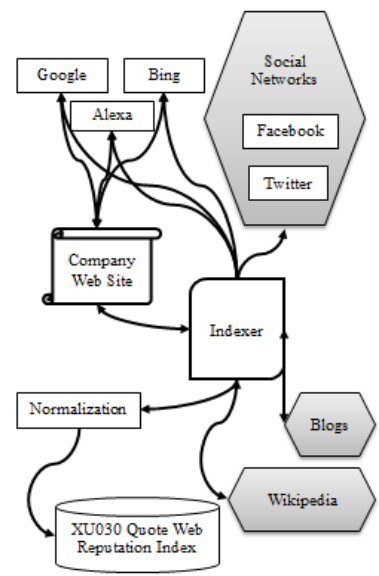

Fig. 1. Data flow diagram of the study.

As it is demonstrated on Fig. 1, the web indicators of a web page of all the companies are gathered from the internet resources. In this study, 17 different parameters are gathered from 11 different sources. After the collection of values by indexer in figure 1, the parameters are normalized to get combined into a single index. The normalized index values are considered as the web reputation index of the XU030 quotes.

## II. Background

Information and communication technology (ICT) has a great impact on firm specific organizational capability. Despite the increasing capabilities correlated with the ICT, scholar are still trying to model the linkage between ICT and financial performance of the firms [4]. Some researches in the MIS area are trying to correlate the statistical ICT data and the performance of the company [5], while some scholars are trying to model the correlation over the ICT and company capabilities, such as financial operations or accounting [6].

Manuscript received February 1, 2014; revised April 24, 2014.
Sadi Evren Seker and Mehmet Lutfi Arslan are with Department of Business, Istanbul Medeniyet University, Istanbul, Turkey (e-mail: academic@sadievrenseker.com, lutfi.arslan@medeniyet.edu.tr).
Bilal Cankir is with the Department of Business, Kirklareli University, Kirklareli, Turkey (e-mail: bilalcankir@gmail.com).

Also the ICT implementations and improvements on the company's yields a transformation on the companies, further replacing classical operations with technology. The transformation brings out a paradigm shift in almost all of the firm capabilities [7].

One of the major capability effected with the ICT transformation is the reputation of the firm. The new paradigm is named as e-reputation and considered as the reputation built on the ICT related operations [8] like social media marketing, electronic communications (e-mails, web forms, social media messaging, etc.) or electronic B2B and B2C operations.

For example, some scholars concentrate on the e-reputation based on the correlation between the social media, such as twitter and the concept map, such as Galois Lattice [9].

According to Thelwall, it is "the study of web-based content with primarily quantitative methods for social science research goals using techniques that are not specific to one field of study" [10]. First example of this measurement is the "Web Impact Factor" (WIF) developed by Ingwersen and defined as "the number of web pages in a web site receiving links from other web sites, divided by the number of web pages published in the site that are accessible to the crawler" [11].

There are five set of tools of web-o-metric research: link analysis, web citation analysis, search engine evaluation, descriptive studies of the web, and the analysis of Web 2.0 phenomena. While link analysis measures the hyperlinks between web pages, web citation analysis counts how often journal articles are cited. Search engines are used to evaluate the extent of the coverage of the web and the accuracy of the reported results. Descriptive studies include various survey methods like the average web page size, average number and type of meta-tags used, the average use of technologies like Java and JavaScript, the number of users, pages and web servers. Last but not least tool is Web 2.0 applications [12].

As the aim of our paper, we use tools of web-o-metric research like Google page rank, number of visitors, number of pages linking back to the web page or the number of likes on Facebook, in order to create a reputation index [13]. Our intention is to be as simple and usable as possible.

Furthermore we conduct the web-o-metric with other ICT environments, which can affect the reputation of the company, such as social media [14], blogs [15] and wikis [16], [18] ,[19].

### III. REPUTATION EXTRACTION

The reputation extraction is built over 4 major groups, which are social media, web-o-metrics, blogs and wiki statistics. Those 4 major groups have 17 different parameters collected from different independent resources. This chapter explains the details of the parameters.

**Wikipedia Page Views**: This parameter indicates the number of views for the Wikipedia entry of the company. Only one of the top 30 companies in Turkish stock market, does not have Wikipedia entry, so it is considered as 0 page views and rest 29 have the entries. The maximum page view is 12,259 for Turkish Airlines.

**Wikipedia Language Count:** This parameter indicates the number of different language entries. Wikipedia supports 287 different languages and some of the companies have entries in multiple languages. For example, the maximum language entry is for Turkish Airlines with 46 different languages.

**Linked-in Follower Count:** The number of people following company in the Linked-in. Again there are several companies without linked-in company page and they are considered as 0 followers. The highest number of followers is 68,114 for Turkcell, cellular phone operator.

**Hate-marks.** There are some web pages in Turkish, who collects the hate-marks from customers. The customers create a user account with declaring their true identity and they can write about their complaints to the web page. The company representative can connect to those web pages also and answer the complaints. We have also included the number of complaints as a hate-mark and taken into account as negative reputation parameter. Again some of the companies do not have any entry, which most of them are operating B2B and have only a few customers. On the other hand the highest number of hate-mark is 18964 for Garanti Bank.

**Love-marks.** Same web sites, who are collecting the hate-marks are also collecting the love-marks from customers. Users can leave their 'thanks' to the company via the same web pages. Besides the companies without any entry the highest number of love-mark is 822 again for Garanti Bank.

**Has a Facebook Page?** We have checked whether the companies have a Facebook page or not. Fortunately all of the companies have a Facebook page except one so we have removed this metric from our calculations.

**Facebook Like Count**. Facebook is the leading social network with highest number of members around the world. We have collected the number of like counts for each of the companies. The maximum like count is for Turkcell and the count is 2.747.255. The minimum value is 0 for the company without the Facebook page. The average value for the Facebook like count is 273.693 and the reason of high standard deviation can be related to the Facebook campaigns of some companies. For example the highest Facebook like count company is a well known telecom company with the Facebook campaigns.

**Value of the Site**. Some of the independent organizations offers a free agent to calculate the expected value of the web site via the web indicators like Alexa ranking or Google page rank. Most of them are built on the number of visitors and expected click from the visitors to make a valuing. The maximum expected value of a company web page is 621.305 and average value of the company sites in Turkey is 105.724 USD.

**BING Backlinks**. The BING back links are collected from the search engine of Microsoft, Bing. The maximum number of Bing backlinks is 3540 for Akbank and the average number of back links is 137.

**Google Backlinks**. Google backlink count is the number of page sites indexed by the Google crawler. This number is under the effect of two facts. First, the number of pages held on the web site is limited. For example if a web site has only 1 page, the maximum possibility for the Google backlinks is 1. Second, even the web page can hold multiple pages, Google can crawl only a part of the web pages. The maximum number of back links is 3.313.000 for TurkTelekom, while the average is 307.817 for all 30 companies.

**Daily Unique Visitors** is the average number of visitors per day. The daily visitor number can vary from date to date and we have collected the up to date values during the research time. The maximum visitor is 637.285 for Garanti Bank and the average for 30 companies is 62.656.

**Alexa Ranking** is another indicator published by an Amazon owned web site alexa.com. The lesser number means the web page has a higher ranking and the minimum ranking for the web site is 24 in Turkey and highest ranking is 65.836 among the whole Turkish web sites on the Internet. Another parameter is the Alexa global ranking, similar to the Turkish ranking. The lowest global ranking is 1.442 and the average is 570.013 among all the web sites on the Internet.

**Time on Site** is a web indicator to measure the time spending of the users with a time interval of their entrance and exit. The higher time means a higher reputation for the web site and the maximum value of time spent on the web page is about 8 minutes and average is about 4 minutes. These time intervals are also daily, which means the time on site indicator is an average day based time spending on web page for each of the user.

**Facebook Shares** is another indicator that is the count of the shares of the web site of the company. The value is fetched from Facebook and the higher number of shares is considered as a positive indicator for the company web site. Unfortunately the numbers available for public access on Facebook is limited with last month. So the number of shares are only limited with last 30 days. The average share count is 211 and the maximum count is 1969 for Turkcell.

**Tweets** parameter is the count of tweets mentioning about the web site of the company. Again, similar to the Facebook shares, the publicly available tweets are limited. Maximum number of tweets is 276 for Halkbank and the average number is 22.

**Google Trend** is the publicly available trend calculator built on the Google search data. Trends values can be both queried as a time series or as the latest value of the trend. We have also added the google trends values for each of the companies in BIST30 as their brand values. The trend values of the brands vary from 19 to 100 where the 100 is the maximum available and 0 is the minimum possibility of the google trends.

## IV. Normalization

In the normalization phase, the collected web indicator values are normalized via min-max normalization.

$$N_{MinMax}(x) = \frac{x - Min}{Max - Min}$$

The normalized value is calculated by the subtraction of the minimum value of the series from the sample and dividing the subtraction to the distance between minimum and maximum values of the series.

The reason of normalization is getting comparable values for each of the indicators. For example, some of the web parameters have values up to millions while some are only limited to 100. For this reason we need a common scale for all of the parameters and we have implemented the min-max normalization for each of the parameters where the result is between 0 and 1.

Another problem in combining multiple parameters into a single metric is the effect of parameters as positive or negative direction. For example the Alexa ranking of a web site can be considered as a negative directed effect on the combination, since the better reputation comes from smaller rankings. As a solution we have calculated the inverse of these indicators by multiplying with -1. Which means a subtraction in the final decision in fact.

So the total score is calculated with below formula.

$$WRI = \frac{\sum_{O}^{C} N_x - \sum_{C}^{K} N_x}{C}$$

The Web Reputation Index (WRI) is calculated with the summation of negative indicators subtracted from the summation of positive indicators divided by the count of positive indicators "C". The "K" symbol in above formula stands for the total number of indicators which is the summation of positive and negative indicator counts.

Because the summation of positive indicators is always higher than the summation of negative indicators the equation of WRI always gets a positive real number between 0 and 1.

## V. Results

This section holds the details of the normalized index values. The complete list of companies with the index values are placed into the appendix of the paper.

Properties of the data set is given in Table I.

TABLE I: Properties of the Index Values

| | |
|---|---|
| Mean ( μ ) | 0.454 |
| Maximum | 1 |
| Minimum | 0.132 |
| Standard Deviation ( σ ) | 0.214 |
| Total Number of Companies | 30 |

The distribution of the company ICT reputation index is given as a separate figure.

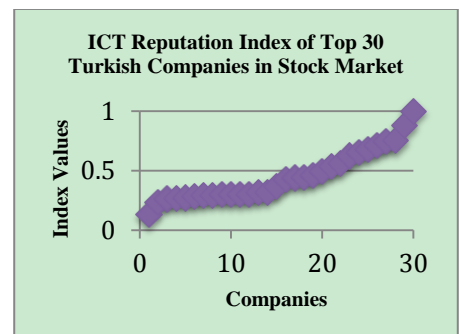

Fig. 2. Statistical distribution of normalized web reputation index.

In Fig. 2, the x-axis holds a unique number for each of the company and all 30 companies are demonstrated on the figure, while the y-axis demonstrates the normalized ICT reputation of the company. The companies are sorted via their ICT reputations and the exact values are given in the appendix.

## VI. Conclusion

Information and communication technologies have an increasing impact on all aspects of the companies. Most of innovative technologies are effecting the success of the companies while some are transforming the business model of the companies.

In this study, the effect of information and communication technology implementation and usage is researched from the view of customer level. A customer can interact the company via the web page of company, social media pages of company, blogs about company or Wikipedia entries of company. We have collected statistics from all these resources for the most prestigious 30 companies, which are quoted to the Istanbul Stock Market (BIST30), and normalized them into a single value.

As a result we have first time publishing the ICT reputation of top 30 companies in Turkish Stock Market via 17 different parameters, which are collected from 11 different independent sources.

The study can be a baseline for further studies about e-reputation, social CRM or new media reputation management.

## Appendix

### Normalized Web Reputation Index For XU030

| Company | WRI |
| --- | --- |
| KOZA MADENCİLİK | 0.132165144 |
| KOZA ALTIN | 0.230420057 |
| IHLAS HOLDİNG | 0.261958342 |
| ERDEMİR | 0.268510418 |
| DOĞAN HOLDİNG | 0.27064328 |
| EMLAK KONUT | 0.282242443 |
| TAV HAVA MEYDANLARI | 0.290815873 |
| ENKA İNŞAAT | 0.292881598 |
| SİŞECAM | 0.29808827 |
| SABANCI HOLDİNG | 0.298591237 |
| MİGROS | 0.300213042 |
| ASELSAN | 0.302485941 |
| KARDEMİR | 0.314336424 |
| KOÇ HOLDİNG | 0.319372511 |
| PETKİM | 0.367171061 |
| TÜPRAŞ | 0.429037531 |
| PEGASUS | 0.437664376 |
| ARÇELİK | 0.441800995 |
| BİM | 0.458792442 |
| BANK ASYA | 0.494751838 |
| VAKIFBANK | 0.536431514 |
| YAPI KREDİ | 0.55969373 |
| TOFAŞ | 0.622762143 |
| TURKCELL | 0.651705972 |
| HALKBANK | 0.679601202 |
| TÜRK TELEKOM | 0.709105825 |
| AKBANK | 0.745626823 |
| THY | 0.751946073 |
| İŞBANK | 0.876582338 |
| GARANTİ | 1 |

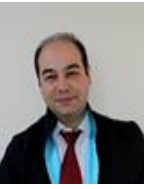**Şadi Evren Şeker** was born in Istanbul in 1979. He has completed his BSc., MSc. and PhD. degrees in computer science major. He also holds an M.A. degree in science technology and society. His main research areas are business intelligence and data mining. During his post-doc study, he has joined data mining research projects in UTDallas. He is currently an Asst. Prof. in Istanbul Medeniyet University, Department of Business. He is an IEEE member and senior member of IEDRC.

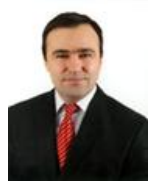**Mehmet Lütfi Arslan** was born in Vezirkopru in 1972. He completed his undergraduate business degree in Marmara University of Istanbul. He obtained his MA and PhD in Social Sciences Institute of Marmara University. During his postgraduate studies, he worked in private sector. He is currently an assistant professor of management and organization in newly founded Istanbul Medeniyet University and studies on management, organizations, entrepreneurship, and human resources management.

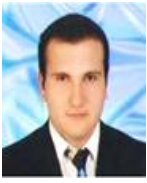**Bilal Çankır** was born in Istanbul in 1985. He applied to a MA program (international quality management) at Marmara University. He finished this school at 2010. He's studied PhD at Istanbul University (management) since 2011 autumn. He is currently working at Kirklareli University Faculty of Economics and Administrative Sciences in Business Administration area since August 2010. He is interested in management, organizations, and organizational behaviour.